# Dust release and tensile strength of the non-volatile layer of cometary nuclei

Yuri Skorov* and Jürgen Blum

Institute for Geophysics and Extraterrestrial Physics,

Technical University of Braunschweig,

Mendelssohn-Str. 3, D-38106 Braunschweig, Germany

[*] Corresponding Author. E-mail address: skorov@mps.mpg.de

Pages: 43

Tables: 0

Figures: 4

**Editorial correspondence to:**

Dr. Yuri Skorov

Technical University of Braunschweig.

Institute for Geophysics and Extraterrestrial Physics,

Mendelssohn-Str. 3, D-38106 Braunschweig, Germany

E-mail address: skorov@mps.mpg.de

**Abstract**

We developed a thermophysical model for cometary nuclei, which is based upon the assumption that comets form by the gravitational instability of an ensemble of dust and ice aggregates. Under this condition, the tensile strength of the ice-free outer layers of a cometary nucleus can be calculated, using the dust-aggregate collision and adhesion model of Weidling et al. (2011). Based on available laboratory data on the gas permeability and thermal conductivity of ice-free dust layers, we derived the temperature and pressure at the dust-ice interface for pure water and pure carbon dioxide ice. Comparison of the vapor pressure below the dust crust with its tensile strength allows the prediction of dust release from cometary surfaces. Our model predicts dust activity for pure $CO_2$ ice and for heliocentric distances of $< \sim 3$ AU, whereas pure $H_2O$ ice cannot explain the dust emission.

# 1. Introduction.

In this paper, we propose a new model to assess the strength of the surface layer of cometary nuclei and investigate the gas release and, thus, the activity of comets. These problems have been studied in numerous cometary publications and different approaches have been applied to estimate the strength of cometary material. We list here only some of them. Greenberg et al. (1995) used observational data of the splitting of comet Shoemaker-Levy 9 to get a quantitative derivation of the tensile strength of the nucleus using a simple microscopic model of intermolecular interactions. Davidsson (2001) and Toth and Lisse (2006) estimated the lower limit of the bulk density and tensile strength from the investigation of tidal splitting and rotational breakup of cometary nuclei. Kührt and Keller (1994) noted that the cohesion between the dust (and ice) particles leads to a tensile strength much larger than the one given by pure gravitational binding. The problem recently received new attention due to the *in situ* experiment executed by the Deep Impact spacecraft. Although the first analysis of the experimental results yielded an extremely low strength of <65 Pa for the cometary crust (A'Hearn et al., 2005), further investigations showed that any strengths from 0 to 12 kPa could fit the observational data (Holsappe and Housen, 2007). There are many other indirect evaluations of the expected strengths of the cometary material and we refer the reader, for example, to Biele at al. (2009) where different approaches are discussed. Here we only note that all assessments, excluding the first interpretation of the Deep Impact results, predict a tensile strength of the order of several hundred or even thousand Pascal. These expectations are in agreement with laboratory measurements of the tensile strength of homogeneous loose packings of micrometer-sized dust particles (Blum and Schräpler, 2004; Blum et al., 2006).

If this supposition is true, one inevitably comes to the conclusion formulated in (Kührt and Keller, 1994), that, when the dust layer has such a cohesive nature, the sublimating gas cannot release dust, because the cohesive strength of the material is several orders of magnitude stronger than the gas pressure underneath the crust. This has the consequence that the cometary activity monotonously diminishes with time and ultimately ceases completely. It seems that the apparent soundness of this conclusion led to the amusing situation that in many papers studying the activity of comets, the dust layer is simply considered as cohesionless. This case was also considered by Kührt and Keller (1994), and later this idea was used, for example, in de Sanctis et al. (2010) and Lasue et al. (2008). This un-physical model assumption allows the easy release of dust particles from the nucleus surface, because the gas pressure exceeds the gravitational pressure of the dust layer, which thereby is able to maintain its activity for a long time. This situation was considered as satisfactory until recently, when new spacecraft flybys resulted in a lot of stunning images as well as spectral and infrared observations of cometary nuclei. We can summarize the recent findings such that ice sublimation takes place underneath a dry, hot, cohesive dust crust and that dust particle emission is correlated with the observed gas activity (see, e.g., A'Hearn et al., 2011). This means that the problem of the disruption of the dust layer by the outflow of gas should be reassessed. The new observations motivated us to revise the problem of dust release by the release of volatile gas molecules on comet nuclei, using a new model of the formation of cometesimals. In Section 2, we present our new approach to the formation of cometesimals and to derivation of the tensile strength of the outer ice-free layers of comet nuclei. In Section 3, we estimate the gas pressure underneath the porous cohesive dust layer over a wide range of parameters, and in Section 4, we discuss the obtained results and constraints in the context of cometary activity.

## 2. A new model of the dust-covered cometary nucleus

In this section, we will describe a new model of the comet nucleus based upon recent developments of formation scenarios for planetesimals. As we will show, the model is full self-consistent and predicts the mechanical strength of the ice-free outer layer of the nucleus. In connection with a heat-transfer and a gas-permeability model it allows the prediction of the activity of comets when approaching the Sun.

## 2.1 Model description

Our model of the formation and evolution of comet nuclei is a multi-stage process whose individual sub-processes are described in the following.

1. **Formation of dust and ice aggregates in the inner and outer solar nebula.** We know from observations (see, e.g., Wilner et al., 2005) and models (see, e.g., Zsom et al., 2010; 2011) that protoplanetary dust rapidly coagulates into (at least) mm-sized dust aggregates. The timescale for this growth is of the order of $10^3$ - $10^4$ years (Zsom et al., 2010; 2011). Obstacles to a further direct growth into planetesimals or cometesimals are the so-called bouncing and fragmentation barriers. Growing dust aggregates reaching the bouncing barrier undergo a (gradual) transition from growth to bouncing (Güttler et al., 2010; Weidling et al., 2011) so that their further gain in mass is stopped at sizes of 0.1-10 mm (depending on the solar-nebula model; see Zsom et al., 2010 for details) and simultaneously get compacted in non-sticking collisions (Weidling et al., 2009). Assuming that further growth beyond the bouncing barrier is possible, dust aggregates will fragment when they reach the fragmentation barrier (Güttler et al., 2010; Beitz et al., 2011; Zsom et al., 2011). Depending on the solar-nebula model, the threshold size for fragmentation falls in the range of millimeters to meters. It is therefore plausible to assume that the minimum and maximum dust-aggregate sizes in the solar nebula were ~0.1 mm and ~1 m, respectively.

2. **Transport of refractory-dust aggregates to the outer solar nebula.** Analysis of Stardust samples collected at comet Wild 2 show that the refractory dust consists high-temperature condensates which can only be formed in the inner solar nebula (McKeegan et al. 2006; Zolensky et al, 2006). Thus, we assume that the dust aggregates, which had been formed in the inner solar nebula, were transported to the cometesimal-forming regions of the protoplanetary disk. During the transport process, we assume that the dust aggregates did not undergo any further growth or fragmentation event.

3. **The dust and ice composition of the outer solar nebula shortly before cometesimal formation.** Shortly before the cometesimals were formed by gravitational instability in an ensemble of dust and ice aggregates (see next point), we assume for simplicity that the number density and size of the dust and ice aggregates floating in the outer solar nebula were identical and that the size distribution of the aggregates was monodisperse. We are aware that these are major restrictions but we believe that the general conclusions of our model are not severely affected by these model assumptions. However, future versions of the model will have to deal with size distributions of dust and ice aggregates as well as with different mass ratios of dust and ice.

4. **Formation of cometesimals by gravitational instability.** The formation of planetesimals and cometesimals were likely caused by a collective gravitational instability of an ensemble of dust and ice aggregates as described, e.g., by Johansen et al. (2007). Following this model, cometary nuclei should then consist of an agglomeration of dust and ice agglomerates with distinct sizes, because neither the impact velocities nor the hydrostatic pressure were high enough to cause considerable compaction and severe homogenization of such km-sized bodies (see points below).

5. **The consolidation of cometesimals.** We conceive the collapse of the dust- and ice-aggregate cloud into a cometesimal as an inward-out process. Dust and ice aggregates close to the center of the collapsing cloud form sticking contacts first. Outer layers subsequently rain down on the forming cometesimal. After a considerable fraction of the final size of the cometesimal has been reached, we assume that the still infalling dust and ice aggregates hit the surface of the growing cometesimal at escape velocity, which is (for not too small cometesimals and not too small dust/ice aggregates) by far larger than the sticking threshold of the aggregates (see Sect. 2.2). Thus, the aggregates will inelastically bounce several times before they will ultimately stick to one another at velocities close to the sticking threshold (see Fig. 1a).

6. **The structure of the cometesimals.** We assume that the process described in the previous point leads to a rather compact packing of the dust and ice aggregates although empirical evidence for this is lacking. Granular matter (with low adhesion force with respect to the gravitational force) typically forms random close packing (RCP) structures. In our model, we assume that the cometesimals formed by the process described above may be slightly less compact and possess structures between random loose packing (RLP) and RCP (see also Sect. 2.2).

7. **Sublimation of the ice aggregates at solar approach.** When the comet approaches the Sun for the first time, the ice aggregates close to the surface evaporate without dragging the dust aggregates along. This is due to the fact that for building up a gas pressure sufficiently high for the release of the dust aggregates, low gas permeability is required. However, close to the surface of the comet nucleus the evaporating gas molecules can almost freely escape (Gundlach et al., 2011a) so that the evaporation of the ice close to the surface of the nucleus should be possible without affecting the dust aggregates.

8. **The two-layer model of comet nuclei.** Following the above model points, the comet nucleus consists in its bulk of a mixture of ice and dust aggregates upon which a layer of ice-free dust aggregates is residing (see Fig. 1b). The morphology of this dust-aggregate layer can be two-fold: (a) The evaporating ice within the ice-dust network and close to the surface of the nucleus does not affect the contacts among the dust aggregates and, thus, it also does not affect the positions of the dust aggregates; the resulting ice-free dust-aggregate layer will then have a volume filling factor half as big as in the original ice-dust matrix, with more spacing in between the dust aggregates. (b) The dust aggregates within the surface layer of the nucleus out of which the ice evaporates are no longer mechanically supported by the underlying aggregates; thus, some of the aggregates collapse to a lower position of the nucleus and stick again to the underlying dust aggregates, but this time their sticking velocity is determined by the minimum of the sticking threshold and the terminal velocity that the aggregates achieve during their free fall.

9. **Thickness of the ice-free dust-aggregate layer.** In our two-layer model of the composition of cometary nuclei, the thickness of the ice-free dust-aggregate layer is a free parameter, along with the dust-aggregate size. Below this refractory dust layer, a pristine ice-dust mixture exists in which the original compositions of the icy and dusty agglomerates have been preserved.

## 2.2 The physical properties of the cometary surface layer.

The most relevant physical property for the determination of the dust activity of comets is the tensile strength. Here, we develop a new derivation of the tensile strength of an *agglomerate of dust aggregates*. The model assumes that dust and ice aggregates of radii $r$ were formed by collisional coagulation in the solar nebula and then were jointly incorporated into a cometesimal of radius $R_c$ (see Sect. 2.1). The individual dust aggregates possess a volume filling factor or packing fraction (i.e. the fraction of the aggregates' volume actually filled with solid dust or ice particles) of $\phi_a$. The volume filling factor of the ice-free dust-aggregate packing structure is $\phi_p$ and the volume filling factor of the entire dust layer is then given by $\phi_l = \phi_a \phi_p$. We assume that the dust and ice aggregates went through the bouncing barrier prior to be embedded into the cometesimal so that, following Weidling et al. (2009) and Zsom et al. (2010), we get $\phi_a \approx 0.4$. If the dust aggregates formed the cometesimal by gravitational instability, the packing density of the (non-sticky) aggregates should be in between the RLP ($\phi_p \approx 0.55$) and the RCP ($\phi_p \approx 0.64$) limit (Onoda and Lininger, 1990), which then yields an approximate value of $\phi_p \approx 0.6$. Thus, we get a total packing fraction of $\phi_t \approx 0.24$ for the bulk of the comet nucleus. Based upon the assumption of position preservation (see Sect. 2.1, point 8a) of the dust aggregates during the evaporation of the ice close to the surface (from which follows that $\phi_p \approx 0.3$ for the dusty surface layer), we get a volume filling factor of the ice free surface layer of $\phi_t \approx 0.12$. When the dust aggregates reconfigure during the ice evaporation, their ultimate packing density will again be close to the RLP or RCP value ($\phi_p \approx 0.6$, see Sect. 2.1, point 8b) so that the total filling factor of the dusty surface layer will be $\phi_t \approx 0.24$.

To derive the tensile strength of a layer of dust aggregates, we apply the model of Weidling et al. (2011) for the surface energy of dust aggregates (their Eq. 5) and use the proportionality between the surface energy $\gamma$ and the tensile strength $T$, i.e. $\gamma \propto T$, so that we get for the effective tensile strength of the dust layer

$$T_{\text{eff}} = T_a \, \phi_p \, A \, / \, A_0 \, , \tag{1}$$

with $A$ and $A_0 = \pi \, r^2$ being the contact area between two dust aggregates and the cross-sectional area of the aggregates, respectively. The intrinsic tensile strength of the aggregates can be estimated by using Eq. 8 of Güttler et al. (2009)

$$\lg(T_a(\phi_a) = 2.8 + 1.4\phi_a \; [\text{Pa}]. \tag{2}$$

With the expected value of $\phi_a \approx 0.4$, we get $T_a = 2300$ Pa, which is the tensile strength of a homogeneous packing of micrometer-sized dust particles. If the dust aggregates composing the dust layers were completely non-cohesive, like particles in a granular medium, the contact area $A$ between two neighboring aggregates would be minimal so that the effective tensile strength would vanish. However, due to mutual collisions during the gravitational collapse phase and attractive inter-particle forces, neighboring dust aggregates exhibit a finite contact area $A$, which will be derived in the following. For this, we use the recent approach by Weidling et al. (2011) who derived the contact area of two colliding dust aggregates (their Eq. 10)

$$A = \pi \left( \frac{15 M^* r^{*2} v^2}{8E} \right)^{2/5},$$ (3)

assuming a Hertzian contact established between the dust aggregates (masses $M_1$, $M_2$, radii $r_1$, $r_2$) in a collision with velocity $v$. Here, $M^*$, $r^*$, and $E$ are the reduced mass of the two dust aggregates (given by $M^{*-1} = M_1^{-1} + M_2^{-1}$), their reduced radius (given by $r^{*-1} = r_1^{-1} + r_2^{-1}$), and Youngs's modulus of the dust aggregate, respectively. For Young's modulus we assume $E = 8100$ Pa (Weidling et al., 2011). In the simplified case of equal-size dust aggregates with masses $m$ and radii $r$, the reduced radius and mass are given by $M^* = M/2$ and $r^* = r/2$, respectively, so that Eq. (3) can be written as

$$A = \pi \left( \frac{15 M r^2 v^2}{64 E} \right)^{2/5}.$$ (4)

All we need to know is the average collision velocity with which the dust aggregates collide (and stick!). As we assume that the cometesimal was formed by gravitational instability rather than by direct collisional growth, the impact velocity of the dust aggregates during the gravitational collapse is approximately given by the escape speed from their final location which is mostly a function of the current comet nucleus radius $R_c$. However, for all practical cases, the escape velocity exceeds the sticking threshold (see Fig. 11 in Güttler et al., 2010) so that the gravitational collapse goes through a succession of bouncing collisions between the dust aggregates on the surface of the cometesimal (see Fig. 1a). Although these bouncing collisions might slightly influence the packing density of the aggregates (see Weidling et al., 2009), they do not contribute to the tensile strength of the final body. Due to the energy loss in each bouncing collisions, the impact velocity of the collapsing dust aggregates decreases until the sticking threshold is reached. Thus, we use the transition velocity from sticking (for low velocities) to bouncing (for high velocities) as $v$ in Eqs. (3) and (4). This threshold velocity was theoretically determined by Güttler et al. (2010) and empirically confirmed (on the 50% sticking probability level) by Weidling et al. (2011). They find for the sticking threshold as a function of dust-aggregate radius

$$\left(\frac{v}{1mm/s}\right) = 0.304\left(\frac{r}{1mm}\right)^{-5/6} \qquad (5)$$

(Eq. 12 in Güttler et al. (2010) and Eq. 8 in Weidling et al. (2011), and assuming a packing density of $\phi_a = 0.4$, and, hence, a mass density of the dust aggregates of 1200 kg/m$^3$). Inserting Eq. (5) into Eqs. (4) and (1) yields

$$T_{eff} = T_l \phi_p \left(\frac{r}{1mm}\right)^{-2/3}, \qquad (6)$$

with $T_l = 1.6$ Pa. For $\phi_p = 0.3$ (case (a) in Sect. 2.1 point 8), we get an effective tensile strength of $T_{eff} = 0.5$ Pa for $r = 1$mm dust aggregates. This is well below earlier estimates of the tensile strength of homogeneous dust layers (see Kührt and Keller, 1994; Blum et al., 2006). In case of reconfiguration of the dust-aggregate layer during the ice evaporation (which then yields again $\phi_p = 0.6$; case (b) in Sect. 2.1 point 8), we have to determine whether the falling dust aggregates hit other dust aggregates deeper within the gravitational potential of the comet nucleus above or below their sticking threshold velocity. The gravitational acceleration on the surface of our model comet is given by

$$a = \frac{4}{3}\pi R_c G \phi_t \rho, \tag{7}$$

with $G$, $\rho$ and $R_c$ being the gravitational constant, the mean material density of the comet, and the radius of the comet nucleus, respectively. The free-fall height of a dust aggregate, which lost its contacts through the evaporating ice, should be a few times its radius. Here, we assume that the dust aggregate is able to fall one aggregate diameter before it hits another dust aggregate. Thus, the collision velocity becomes

$$v = 2a^{1/2}r^{1/2} = 7.3 \times 10^{-4} m/s \left(\frac{Rc}{1km}\right)^{1/2} \left(\frac{r}{1mm}\right)^{1/2}. \tag{8}$$

Here we used a total packing density of $\phi_t = 0.24$ and a mean material density of $\rho = 2000$ kg/m³ of the comet nucleus. A comparison with Eq. (5) shows that dust aggregates larger than about 0.5 mm (for $R_c = 1$km) fall down faster than their sticking threshold so that for these aggregate sizes Eq. (6) still describes the resulting tensile strength (however, with $\phi_p = 0.6$). Smaller dust aggregates are, according to Eq. (8), less bound after their re-organization so that Eq. (6) has to be replaced for those dust aggregates which went through the process described above by

$$T_{eff} = T_l \phi_p \left(\frac{R_c}{1km}\right)^{2/5} \left(\frac{r}{1mm}\right)^{2/5}, \tag{9}$$

with $T_l = 3.3$ Pa and $\phi_p = 0.6$.

In the former case (Eq. 6), the tensile strength of a layer of porous dust aggregates continuously decreases with increasing aggregate size and obtains a value of ~0.5 Pa for mm-sized dust aggregates, which is more than three orders of magnitude below the tensile strength of a homogeneous (i.e. not hierarchically grown) dust layer, which is given by $T_t = 1000$ Pa ($\phi_a = 0.15$) … 6300 Pa ($\phi_a = 0.66$) (Blum et al., 2006) (see also Eq. (2) for the tensile strength of a homogeneous particle packing with random packing faction). In the latter case (Eq. 9), the tensile strength for sub-mm sized dust aggregates, which underwent a restructuring event, is even smaller and increases with increasing aggregate size.

What remains to do is to compare the tensile strengths determined by Eqs. (6) and (9) as a function of the dust-aggregate radius with the gas pressure (vapor pressure) under the dust layer. Emission of dust can only occur if the gas pressure is higher than the tensile strength.

One has to mind that our model does not only predict the tensile strength of the ice-free dust layer of a comet nucleus but also the tensile strength of the bulk nucleus. Here, Eq. (6) can also be applied with $\phi_p = 0.6$ and a material-dependent cohesive strength, which can considerably increase the value of $T_l$. Under the above assumption that the tensile strength is proportional to the surface energy of the particle material, the tensile strength of the comet nucleus as a whole can exceed that of the ice-free surface layer by at least a factor of ten, due to an increased surface energy of water ice (Gundlach et al., 2011b). On top of that, the gravitational strength of larger bodies also needs to be taken into account.

## 2.3 Mass and energy balance in the surface layers.

It is well known that the formation of a porous dust layer on the surface of the cometary nucleus generally leads to an increase in temperature of the ice surface and to a corresponding increase of the gas pressure below the nonvolatile dust layer. This effect is clearly observed in laboratory experiments (Benkhoff et al., 1995; Koemle et al., 1990) and has a simple theoretical explanation.

Indeed, if the ice temperature is fixed, the porous dust layer greatly reduces the effective rate of sublimation for all regimes of gas diffusion (from Knudsen diffusion to a continuous flow). For the collisionless gas flow (Knudsen regime), the mass flow through a porous crust is given by

$$\Psi_K = F(r_d, L, \phi)\left(\frac{P_s(T_i)}{\sqrt{T_i}}\right),$$ (10)

where $F(r_d, L, \phi)$ is a model-dependent function describing the geometrical structure of the porous medium, characterizing by the dust-layer thickness $L$, the effective pore radius $r_d$ and the filling factor $\phi$ of the homogenous dust layer. $P_s$ is the equilibrium vapor pressure for the temperature $T_i$ at the surface of the ice. In the simplest case where the porous medium is constructed by cylindrical straight channels (capillaries), Eq. (10) transforms into Clausing's formula (Steiner, 1990; Skorov et al., 2011), with

$$F(r_d, L, \phi) = (1-\phi)\frac{20 + 8(L/r_d)}{20 + 19(L/r_d) + 3(L/r_d)^2}.$$ (11)

The obvious simplification of both above formulae is a treatment of pores as a bundle of straight cylindrical isolated channels. Therefore, in this paper we use for $F(r_d, L_d, \phi)$ the empirical dependence of the permeability of a randomly packed porous dust layer on the filling factor, the thickness of the layer and the grain (here dust aggregate) size (Eq. 19 in Gundlach et al., 2011a)

$$F(r, L, \phi) = \left(1 + L/2r_d b\right)^{-1},$$ (12)

where $b$ denotes the thickness in particle diameters at which the permeability drops to 50 percent. Gundlach et al. (2011a) derived with two experimental setups two values for this half-thickness, $b = 7.31$ and $b = 6.54$ particle diameters, respectively. In all cases it is readily seen that, as the thickness of a dust crust increases, a considerable quenching of the emergent gas flux occurs for a fixed temperature $T_i$.

The presented consideration of gas flow abatement is valid if the *surface temperature of the ice* is fixed. However, for cometary nuclei, the *incoming energy flux* is fixed, because the characteristic time scales for gas diffusion and heat conduction are much smaller than the characteristic time of irradiation changes on the cometary surface. Thus, in the real case one has to consider both energy balances, i.e. on the top (index $s$ - cometary surface) and on the bottom (index $i$ - dust-ice boundary) of the dust layer, in order to evaluate the resulting gas flow through the porous dust layer and the resulting gas pressure above the sublimating surface. Assuming the heat conductivities of the dry dust and the ice-dust mixture as well as the sublimation energy of the ice to be constant, we get a system of nonlinear algebraic equations, namely

$$J = \varepsilon \sigma T_s^4 - \lambda_d \frac{T_s - T_i}{L}, \tag{13}$$

$$\lambda_d \frac{T_s - T_i}{L} = \lambda_i \frac{T_i - T_e}{\Delta S} + \frac{Q\Psi_K}{m}. \tag{14}$$

Here, $J$ is the incoming radiation energy absorbed at the surface of the cometary nucleus, $\varepsilon$ is the infrared emissivity, $\sigma$ is the Stefan–Boltzmann constant, $T_s$ and $T_i$ are the temperatures at the top and the bottom of the dust layer, $\lambda_d$ and $\lambda_i$ are the heat conductivities of the dust layer and the ice-dust mixture, $Q$ is the heat of sublimation per molecule, $\Psi_K$ is the mass flux, $T_e$ is the so-called equilibrium temperature of the nucleus core, and $\Delta S$ is the effective orbital skin depth, respectively. The two latter quantities can be calculated by the method presented by Klinger (1981). Equations (13) and (14) compose a closed system describing the energy balance in a model with the following free parameters: $J, \varepsilon, L, r_d, \lambda_d, \lambda_i, T_e, \Delta S,$ and mass flux $\Psi_K$.

## 3. Temperature and gas pressure underneath the porous dust layer.

We evaluated the model presented above (Eqs. 13 and 14) together with the empirical gas permeability model (Eq. 12) for two heliocentric distances ($R$=1AU and $R$=3AU, respectively) and for pure water ice as well as for pure carbon dioxide ice, respectively, each covered by a dust layer consisting of dust aggregates with sizes $r$.

Obviously, the model assumption about the pure ice is a strong idealization of the real nature of cometary ices. Apparently, the cometary material is a complex heterogeneous mixture of volatile and nonvolatile components. Both the composition and structure of the material can vary considerably from place to place. Nevertheless, considered simplified model makes sense for two reasons. First, at present a theoretical model of the sublimation of a multi-component ice in cometary environments is not developed properly. The only one attempt of a simplified theoretical analysis of a multicomponent sublimation is presented in (Shul'man, 1987). Furthermore the experimental data about multicomponent ice sublimation at low temperature in a vacuum are very limited. In such a situation, we believe that a consideration of a ice mixture would be premature and entirely speculative. At the same time, consideration of idealized one-component model of the ice allows us to estimate expected range of variation of the studied physical characteristics. Obviously, the rate of sublimation (and hence the corresponding pressure of sublimating gas) of any mixture of water and carbon dioxide will lie between these extreme cases of pure ice. This is the second justification for the simplified model considered here.

The gas pressure at the sublimating $H_2O$ ice surface and the sublimating $CO_2$ ice surface is calculated. $H_2O$ ice is traditionally treated as the major volatile component of a cometary nucleus, whereas $CO_2$ ice recently turned out to be the focus of attention (A'Hearn et al, 2011). For simplicity, we neglect the presence of dust aggregates within the ice layers and assume that their influence on the sublimation process is small. The heat capacity, the heat conductivity, the latent heat of sublimation and the expression for the saturation pressure are taken from Fanale and Salvail (1984). The emissivity $\varepsilon$ equals (1-albedo), according to Kirchhoff's law, with an albedo of 0.04, and for the equilibrium temperature and skin depth we chose $T_e$ = 90K and $\Delta S$ =10m, respectively. It is worth noting that the simulation results are quite insensitive to the specific values of two latter parameters. The effective thermal conductivity of the dry porous dust layer used in this paper is $\lambda_d$ = 0.02 W/(m·K), in agreement with measurements by Krause et al. (2011), who determined the heat conductivity of hierarchical dust samples (i.e. agglomerates of dust aggregates, see Sect. 2). The simulation results are shown in Fig. 2 (for water ice) and Fig. 3 (for carbon dioxide ice) which depict in the top row the temperatures of the sublimating ice surfaces as a function of the thickness of the dust-aggregate layer for several sizes of dust aggregates and the corresponding saturation pressure in the bottom panels for $R$=1AU (left panels) and $R$=3AU (right panels), respectively. We consider although two structural models of a porous dust layer: with the filling factor $\phi_p$ = 0.3 (case (a) in Sect. 2.1 point 8) and $\phi_p$ = 0.6 (case (b) in Sect. 2.1 point 8).

When the incoming energy flux is fixed, the energy balance at the sublimating boundary occurs at a higher temperature (in comparison with the temperature of bare irradiated ice) and hence at a higher gas pressure if a dust layer is present, due to the so-called "cooking effect". This effect was observed in laboratory experiments (Benkhoff et al., 1995; Koemle et al., 1990) and reproduced in numerical models (Koemle and Steiner, 1992; Skorov et al., 1999). Because the saturation pressure is an exponentially growing function of temperature, we consider first, how the ice temperature varies with increasing dust-layer thickness and dust-aggregate size. It is clear from Figs. 2 and 3 that the ice temperature $T_i$ is more sensitive to changes of the dust-aggregate size than to layer-thickness variations. The ice surface temperature is visibly higher for small pore radii (i.e. for small dust aggregates sizes), but the maximum temperature is still smaller than the temperature of ice melting in all examined cases.

As might be expected the reduction of porosity of the dust layer leads to a decrease in the effective sublimation, which in turn leads to an increase in both temperature and gas pressure. Because the latter is an exponential function of temperature, even a relatively small increase of temperature induces a pressure increase in several times. Thus one can conclude that local *structural* inhomogeneities may lead to local dust layer demolition and gas outburst.

It is worth noting that for an extended porous layer the difference between the surface temperature $T_s$ and the temperature of the sublimating ice $T_i$ does not depend on the thickness of the crust $L$. Indeed, in this case theory (Eq. 11) and experiments (Gundlach et al., 2011a) predict that the relative permeability of the layer is inversely proportional to its thickness. Substituting this dependence for $\Psi$ into Eq. (14) and expressing the energy conservation at the boundary between dust crust and sublimating ice, one gets a relation connecting $T_s$ with $T_i$ and not containing $L$. For a Hagen–Poiseuille flow an analogous result was found by Shul'man (1987). This fact leads to an interesting conclusion: because the value of $T_s$ never exceeds the corresponding black body temperature (for $\varepsilon=1$), there is an upper limit for $T_i$ and, consequently, for the maximum gas pressure under a thick dust crust. This qualitative analysis is in agreement with the simulation results for the saturation pressure presented in Figs. 2 and 3 (low panels). We see that the pressure is more sensitive to the effective pore size than to the layer thickness for not too thin dust layers.

It is interesting to compare the results obtained for water ice and carbon dioxide ice. For fixed ice temperature, the saturation vapor pressure of $CO_2$ is much higher than the corresponding value for $H_2O$; the typical difference is 5-6 orders of magnitude. Consequentially, the evaporation rate of $CO_2$ ice is much higher than the evaporation rate of $H_2O$ ice, which results in a much more efficient cooling of carbon dioxide ice, due to the latent heat of evaporation. In our model, this means that the energy equilibrium at the dust-ice interface is achieved at a lower temperature for $CO_2$ ice if the incoming energy is fixed. Thus, for $R$=1AU the maximum temperature of the water ice is ~100 K above the temperature of the $CO_2$ ice. In turn, this leads to an equilibrium gas pressure above the $CO_2$ ice only a factor of a few higher than the corresponding gas pressure above the water ice. Thus, we conclude from our model that the destructive power of *pure* carbon dioxide, which has been speculated in cometary society to be a "super volatile" and responsible for dust lifting, is comparable with that of *pure* water ice. This picture might, however, change if intimate mixtures of $H_2O$ and $CO_2$ ice are assumed. This will be the subject of further studies.

Finally we compare the calculated gas pressures above the ice surface covered by a porous dust crust with the strength of this crust estimated by Eqs. 6 and 9.

In Fig. 4 we plot the resulting strength of the crust as a function of aggregate size (bold dashed lines) together with the gas pressure of water vapor for different sizes of dust aggregates, for two heliocentric distances (panel A for 1AU and panel B for 3AU) and different thicknesses of the dust crust (triangles for $L$=$10^{-3}$ m, diamonds for $L$=$10^{-2}$ m, squares for $L$=$3 \cdot 10^{-2}$ m, and circles for $L$=$5 \cdot 10^{-2}$ m). Solid curves are used for the filling factor $\phi_p$ = 0.6 (case (b) in Sect. 2.1 point 8), dashed curves - for $\phi_p$ = 0.3 (case (a) in Sect. 2.1 point 8).

If the dust residual retains its original structure after evacuation of all volatiles ($\phi_p$ = 0.3), the effective strength is determined by Eq. 6. For all examined cases, the pressure of water vapor is below the strength of the dust crust. The pressure always drops with increasing aggregate radius and this trend is insensitive to the thickness of the dust layer if the thickness is large enough ($L > 1$cm). At a distance of $R = 3$AU the gas pressure is less than ten per cent of the crust strength, whereas at $R = 1$ AU gas pressure is only a factor of 2-3 below the strength. One can expect that if an accumulation of fine grains between larger coarse dust aggregates exists, which does not considerably affect the tensile strength, the permeability of the dust layer will decrease so that the gas pressure can exceed the crustal strength.

If the structure of dust residual is modified after evacuation of all volatiles ($\phi_p = 0.6$), the effective strength is determined by Eqs. 6 and 9. At a distance of $R = 1$AU for all aggregates with size less than about 10-4 m pressure is visibly above the layer strength and this difference is 1-1.5 orders of magnitude. We conclude that the compaction of depleted dust layer results in its mechanical instability and disruption even in the case of pure water ice sublimation. At a distance of $R = 3$AU the layer above water ice remains stable for all aggregates except the smallest. It means that fine dust can be accumulated at the surface.

For $CO_2$ ice, the situation is quite different (Fig. 5). If the structure of dust residual is unmodified after evacuation of volatiles ($\phi_p = 0.3$), we see that at $R = 3$ AU for the layer with higher permeability ($b = 7.31$) the tensile strength dominates the gas pressure in all cases. The additional calculations shown that the gas pressure can exceed slightly the tensile strength only for the layer with lower permeability ($b = 6.54$), for the smallest dust aggregates and the thickest dust layers. At $R = 1$ AU and both values of $b$, however, the underlying gas pressure can expel the dust crust for almost all combinations of dust-aggregate size and crust thickness, except for the smallest aggregate sizes and for very thin dust layers. For $b = 7.31$ the gas pressure slightly exceeds the tensile strength for the dust-aggregate sizes smaller than ~1 mm and is comparable to the crustal strength for larger dust-aggregates.

If after evacuation of volatiles the dust layer became more compact ($\phi_p = 0.6$), at $R$ = 3AU the state for CO2 ice is similar to the state considered at $R = 1$AU for water ice: for all aggregates with size less than about 10-4 m pressure is visibly above the layer strength (except for thin dust layer (marked by triangles) formed by the small aggregates) and this difference is 1-1.5 orders of magnitude. At $R = 1$AU as before for the unmodified layer, the underlying gas pressure expels the dust crust for almost all combinations of dust-aggregate size and crust thickness, except for the biggest aggregate sizes and for thick dust layers (marked by squares and circles). For the model with the lower layer permeability (b=6.54) gas pressure is always higher the strength.

As it is unclear which value for the parameter $b$ should be chosen, we cannot make any definite predictions with the model at this point. However, the fact that our model results are extremely sensitive to $b$ and that the actual value of $b$ seems crucial for the occurrence of a crust-destroying activity at high solar distances, shows that our model might be close to a realistic description of the dust-crust evolution of cometary nuclei: a much higher value of $b$ will reduce the sub-crustal pressure so much that the crust can never be removed, whereas a much lower value of $b$ will not even allow the formation of a thin dust crust.

## 4. Discussion.

Before proceeding to discuss the obtained results, we consider the main model simplifications. The first and most significant simplification in our model is the assumption that the dust aggregates have the same size and shape, i.e. we have investigated a monodisperse system of dust aggregates. This simplification is important for the estimation of the strength of the dust crust as well as for the calculation of the surface ice temperature and, consequently, the gas pressure underneath the porous dust layer. In this paper, the significance of this simplification is partially offset by our studying a wide range of dust-aggregate sizes and by getting a qualitatively similar behavior of temperature and pressure in most cases. However, it is highly desirable to exclude this simplification in future investigations in which we will develop models for polydisperse media.

The next model restriction derives from the fixed heat conductivity of the dust crust. Although we use experimental values available from proper laboratory experiments (Krause et al., 2011), the environmental conditions and the microscopic structure of the dust aggregates can influence the thermal conductivity. A model for the thermal conductivity of hierarchical dust-aggregate assemblages is highly desirable and should be used in future studies. At small heliocentric distances, the dust temperature may be so high, that radiative heat transfer becomes comparable to or even exceeds the heat transfer trough the solid dust-particle matrix. This effect is particularly important for bigger dust aggregates, because the radiative conductivity is proportional to effective pore size.

The last model simplification is the physical description of icy component. We tested only *pure* ices, whereas cometary ice is a mixture of different volatiles and dust. However, this assumption seems reasonable at the first step of investigation, because at the moment we have no reliable quantitative information about the microscopic as well as the macroscopic organization of cometary ice. Moreover, even the sublimation energy and evaporation rate for multicomponent ice should be first carefully evaluated in laboratory experiments. As it is easy to see from the results presented above, the temperature of water ice can be higher than the corresponding temperature of carbon dioxide ice by about one hundred degrees (for the same set of model parameters). This means that any examination of macroscopic heterogeneous mixtures of $H_2O$ and $CO_2$ ices (assuming that there are only microscopic pure ice particles) plus dust requires a much more sophisticated multidimensional model of the heat transfer in order to accurately simulate the distribution of ice temperatures and the resulting gas pressures.

We now proceed to the discussion of the results presented above. Let us first consider the dependence of the temperature and the gas pressure on the dust-crust thickness and dust-aggregate size. In the (as we believe unrealistic) case of relatively small aggregates ($r \leq 10^{-4}$ m), both for water and carbon dioxide ice, temperatures and, hence, gas pressures generally increase when the thickness of the dust layer is growing (for fixed dust-aggregate size). This increase is obviously determined by the reduction of the gas permeability of the dust crust and, in turn, by the reduction of the effective sublimation. In the cases in which we observe the ice temperature to reach a horizontal asymptotic value for thicker crusts, the gas pressure reaches a maximum value that is a function of dust-aggregate size only. This effect was discussed above. It is interesting to note that for the biggest dust aggregates, the trend is opposite: temperature and pressure decrease slightly with growing crust thickness. This is explained by the fact that for small crust thicknesses the permeability of the crust is still high ($L/r$ < 4-5), whereas the thickness of the crust is already large enough to considerably reduce the incoming heat flux. For a fixed crust thickness, the gas pressure always decreases when the dust-aggregate size increases. It is interested to note that the slopes of the pressure and tensile-strength curves are almost identical for $H_2O$ (see Fig. 4), whereas for $CO_2$ the slope of the decrease of tensile-strength with increasing dust-aggregate size is deeper than the pressure decrease (see Fig. 5). This means that for $H_2O$ an instability of the dust crust cannot be reached by larger dust aggregates so that the effect of dust-layer desorption is more or less independent of the dust-aggregate size (for monodisperse dust aggregates). For $CO_2$ an instability of the dust crust can either be reached with smaller dust aggregates or with thicker dust layers.

Now we consider in more detail how the gas pressure is varying when the crust thickness increases. It is easy to see that the pressure of the water vapor remains below the estimated strength of the dust crust in all examined cases for loose dust layers (Fig. 4). At $R = 3$ AU, the crust strength even exceeds the corresponding water-vapor pressure by more than a factor of ten. This means that as long as there is any gas activity, the dust accumulates on the surface and the thickness of the crust grows. Although the gas pressure generally increases during this process, this pressure increase is not sufficient to lift the dust crust so that the dust layer is stable. At $R = 1$ AU, the situation qualitatively remains the same, but the ratio of the crustal strength to the maximum gas pressure decreases to about two. Thus, one can expect that the dust crust above *pure* water ice can be removed only if the incoming energy flux is large enough (more than one solar constant).

The other process that can destroy a dust layer is its compaction (case (b) in Sect. 2.1 point 8). For this case at $R = 1$AU gas pressure exceeds the layer strength even above the pure water ice if the layer is constructed by fine dust aggregates (size is smaller than hundred of microns), whereas the big particles are accumulated on the surface.

Fortunately, the situation is more encouraging in the case of $CO_2$ ice (Figs. 5). At $R = 3$AU and at the beginning of crust formation (i.e. for a small crust thickness), the gas pressure is smaller than the crust strength. When the thickness of the dust crust increases due to the onset of the ice evaporation the gas pressure can exceed the crust strength of layer with decreased permeability and, hence, the crust is being removed. As a result, the gas pressure drops again and the growth process of the dust crust can start all over again. Thus, we can expect a repetitive process of growing and removing of the dust layer on top of the ice.

It is interesting to note that in the case of large dust aggregates (with sizes bigger than several millimeters) the gas pressure remains below the strength curve and a permanent dust crust can be formed. However, as the comet approaches the Sun, the gas pressure increases (Fig. 5b), the gas pressure exceeds the crustal strength even for large dust aggregates and the dust crust gets removed. One can see that with our current model parameters for $R = 1$ AU and pure CO2, there is no to build an extended dust crust, because the gas pressure underneath such a dust layer would be several times higher than the corresponding mechanical strength of the crust.

We should emphasize again that the *pure* ices examined in this paper are highly idealized cases and model simplifications. However, one can expect that for mixed ices the general trends will be the same and the resulting gas pressure will probably be between the extreme values evaluated in this paper. This means that the main features found in this study, namely i) formation and growth of a dust crust at large heliocentric distances, ii) removal of the dust crust for smaller heliocentric distances, due to a fast increase of gas pressure underneath the crust, and iii) formation of the next generation of the dust layer, will be valid also for mixed ice.

Generally speaking, the results listed above can be found in various cometary publications about the formation and stability of dust mantles on cometary nuclei. Different aspects of this problem were first discussed by Shul'man (1987), Rickman et al. (1990) and Kührt and Keller (1994) and were later analyzed, for example, by Rosenberg and Prialnik (2009) and de Sanctis, et al. (2010). The main difference of our work and its advantagesover previous approaches are related to two important features: a) in this paper for the first time a model describing the microscopic structure of cometary material in accordance with modern models of the Solar System formation is presented, b) the major model characteristics (such as porosity, permeability, heat conductivity, strength) are determined in a consistent manner and based on experimental results. The latterallows us to pass from qualitative speculations to quantitative analysis of linkages and roles of different model parameters.

## 5. Conclusions.

In this paper, we present the first consistent model of the formation and removal of a porous dust crust on cometary nuclei, based on the evaluation of the tensile strength and on the energy transfer within an ice-free surface dust layer on top of pure water or carbon dioxide ice. The strength evaluation is obtained both from theoretical expectations and from laboratory experiments. The heat transfer model also utilizes results retrieved from laboratory experiments. At different heliocentric distances, we calculated the gas pressure underneath a porous dust crust of different thickness formed by dust aggregates of various sizes. We found that:

1) The hierarchic structure of the dust aggregates, which follows from the formation of cometesimals by gravitational instability, dramatically reduces the tensile strength of the porous dust crust formed by such grains.

2) The general existence of a dust crust leads to an increase in the temperature at the ice-dust interface and, hence, to an increased gas pressure.

3) The pressure of pure water ice generally is not sufficient to destroy the dust crust.

4) The pressure of pure carbon dioxide ice can exceed the tensile strength of the crust even at $R > 1$ AU.

5) The structural modification of the dust layer after evacuation of volatiles leads to decrease of its permeability and to significant increase of gas pressure, which can exceed the strength of this depleted layer.

6) Because generally the gas pressure increases with increasing crust thickness, the formation and destruction sequence of the dust crust can be repetitive. As this repetitive event is not synchronized over the comet surface, our model predicts that only a small part of the total surface exhibits pure ice on the surface, in agreement with observations (Sunshine et al., 2006), as a major part of the cycle is consumed by the build-up of a sufficiently thick dust layer.

**References.**

A'Hearn, Michael F.; Belton, Michael J. S.; Delamere, W. Alan; Feaga, Lori M.; Hampton, Donald; Kissel, Jochen; Klaasen, Kenneth P.; McFadden, Lucy A.; Meech, Karen J.; Melosh, H. Jay; Schultz, Peter H.; Sunshine, Jessica M.; Thomas, Peter C.; Veverka, Joseph; Wellnitz, Dennis D.; Yeomans, Donald K.; Besse, Sebastien; Bodewits, Dennis; Bowling, Timothy J.; Carcich, Brian T.; Collins, Steven M.; Farnham, Tony L.; Groussin, Olivier; Hermalyn, Brendan; Kelley, Michael S.; Li, Jian-Yang; Lindler, Don J.; Lisse, Carey M.; McLaughlin, Stephanie A.; Merlin, Frédéric; Protopapa, Silvia; Richardson, James E.; Williams, Jade L. 2011. EPOXI at Comet Hartley 2. Science, 332, 1396-1400.

A'Hearn, M. F.; Belton, M. J. S.; Delamere, W. A.; Kissel, J.; Klaasen, K. P.; McFadden, L. A.; Meech, K. J.; Melosh, H. J.; Schultz, P. H.; Sunshine, J. M.; Thomas, P. C.; Veverka, J.; Yeomans, D. K.; Baca, M. W.; Busko, I.; Crockett, C. J.; Collins, S. M.; Desnoyer, M.; Eberhardy, C. A.; Ernst, C. M.; Farnham, T. L.; Feaga, L.; Groussin, O.; Hampton, D.; Ipatov, S. I.; Li, J.-Y.; Lindler, D.; Lisse, C. M.; Mastrodemos, N.; Owen, W. M.; Richardson, J. E.; Wellnitz, D. D.; White, R. L. 2005. Deep Impact: Excavating Comet Tempel 1, Science, 310, 5746, 258-264.

Beitz, E., Güttler, C., Blum, J., Meisner, T., Teiser, J., Wurm, G. 2011. Low-velocity Collisions of Centimeter-sized Dust Aggregates. Astrophysical Journal 736, article id. 34.

Benkhoff, J., Seidensticker, K. J., Seiferlin, K., Spohn, T. 1995. Energy analysis of porous water ice under space-simulated conditions: results from the KOSI-8 experiment. Planetary and Space Science 43, 353-361.

Biele, J.; Ulamec, S.; Richter, L.; Knollenberg, J.; Kührt, E.; Möhlmann, D. 2009. The putative mechanical strength of comet surface material applied to landing on a comet, Acta Astronautica, 65, 7-8, 1168-1178.

Blum, J., Schräpler, R., Davidsson, B. J. R., Trigo-Rodriguez, J. M. 2006. The Astrophysical Journal, 652, 1768–1781.

Blum, J., Schräpler, R. 2004. Structure and Mechanical Properties of High-Porosity Macroscopic Agglomerates Formed by Random Ballistic Deposition. Physical Review Letters 93, 115503.

Davidsson, B. J. R. 2001. Tidal Splitting and Rotational Breakup of Solid Biaxial Ellipsoids, Icarus, 149, 2, 375-383.

de Sanctis, M. C.; Lasue, J.; Capria, M. T.; Magni, G.; Turrini, D.; Coradini, A. 2010. Shape and obliquity effects on the thermal evolution of the Rosetta target 67P/Churyumov-Gerasimenko cometary nucleus, Icarus, 207, 1, 341-358.

de Sanctis, M. C.; Lasue, J.; Capria, M. T. 2010. Seasonal Effects on Comet Nuclei Evolution: Activity, Internal Structure, and Dust Mantle Formation. The Astronomical Journal, 140, 1-13.

Fanale, F.P., Salvail, J.R. 1984. An idealized short-period comet model - Surface insolation, $H_2O$ flux, dust flux, and mantle evolution. Icarus 60, 476-511.

Greenberg, J. M.; Mizutani, H.; Yamamoto, T. 1995. A new derivation of the tensile strength of cometary nuclei: Application to comet Shoemaker-Levy 9, Astronomy and Astrophysics, 295, 2, L35-L38.

Gundlach, B., Skorov,Y.V., Blum, J. 2011a. Outgassing of icy bodies in the Solar System - I. The sublimation of hexagonal water ice through dust layers. Icarus, 213, 710-719.

Gundlach, B., Kilias, S., Beitz, E., Blum, J. 2011b. Micrometer-sized ice particles for planetary-science experiments – I. Preparation, critical rolling friction force, and specific surface energy. Icarus, 214, 717-723.

Güttler, C., Blum, J., Zsom, A., Ormel, C. W., & Dullemond, C. P. 2010, The outcome of protoplanetary dust growth: pebbles, boulders, or planetesimals?. I. Mapping the zoo of laboratory collision experiments. Astronomy and Astrophysics, 513, A56.

Güttler, C., Krause, M., Geretshauser, R. J., Speith, R., & Blum, J. 2009. The Physics of Protoplanetesimal Dust Agglomerates. IV. Toward a Dynamical Collision Model. Astrophysical Journal, 701, 130-141.

Holsapple, K. A.; Housen, K. R. 2007. A crater and its ejecta: An interpretation of Deep Impact, Icarus, 191, 2, 586-597.

Johansen, A.; Oishi, J. S.; Mac Low, M.-M.; Klahr, H.; Henning, T.; Youdin, A. 2007. Rapid planetesimal formation in turbulent circumstellar disks. Nature, 448, 1022-1025.

Klinger, J. 1981. Some consequences of a phase transition of water ice on the heat balance of comet nuclei. Icarus, 47, 320-324.

Koemle, N.I., Steiner, G., Seidensticker, K.J., Kochan, H., Thomas, H., Thiel, K., Baguhl, M., Hoeppner, B. 1990. Temperature evolution and vapour pressure build-up in porous ices. Planetary and Space Science, 40, 1311-1323.

Koemle, N.I., Steiner, G. 1992. Temperature evolution of porous ice samples covered by a dust mantle. Icarus, 96, 204-212.

Krause, M., Blum, J., Skorov, Y.V., Trieloff, M. 2011. Thermal conductivity measurements of porous dust aggregates: I. Technique, model and first results. Icarus, 214, 286-296.

Kührt, E.; Keller, H.U. 1994. The formation of cometary surface crusts, Icarus, 109, 121–132.

Lasue, J.; de Sanctis, M. C.; Coradini, A.; Magni, G.; Capria, M. T.; Turrini, D.; Levasseur-Regourd, A. C. 2008. Quasi-3-D model to describe topographic effects on non-spherical comet nucleus evolution, Planetary and Space Science, 56, 15, 1977-1991.

McKeegan, K. D.; Aléon, J.; Bradley, J.; Brownlee, D.; Busemann, H.; Butterworth, A.; Chaussidon, M.; Fallon, S.; Floss, C.; Gilmour, J.; Gounelle, M.; Graham, G.; Guan, Yu.; Heck, Ph. R.; Hoppe, P.; Hutcheon, , I. D.; Huth, J.; Ishii, H.; Ito, M.; Jacobsen, S. B.; Kearsley, A.; Leshin, L. A.; Liu, , M.-C.; Lyon, I.; Marhas, K.; Marty, B.; Matrajt, G.; Meibom, A.; Messenger, S.; Mostefaoui, S.; Mukhopadhyay, S.; Nakamura-Messenger, K.; Nittler, L.; Palma, R.; Pepin, R. O.; Papanastassiou, D. A.; Robert, F.; Schlutter, D.; Snead, Ch. J.; Stadermann, F. J.; Stroud, R.; Tsou, P.; Westphal, A.; Young, E. D.; Ziegler, K.; Zimmermann, L.; Zinner, E. 2006. Isotopic Compositions of Cometary Matter Returned by Stardust. Science 314, 1724-1728.

Onoda, G. Y., Liniger, E.G. 1990. Random loose packings of uniform spheres and the dilatancy onset. Physical Review Letters, 64, 2727-2730.

Poppe, T.; Blum, J.; Henning, T. 2000. Analogous Experiments on the Stickiness of Micron-sized Preplanetary Dust. Astrophysical Journal, 533, 1, 454-471.

Rosenberg, E. D.; Prialnik, D. 2009. Fully 3-dimensional calculations or dust mantle formation for a model of Comet 67P/Churyumov-Gerasimenko. Icarus, 201, 740-749.

Shul'man, L.M. 1987. Cometary nuclei. Moscow, Nauka, 230 p.

Skorov, Y.V.,Kömle, N.I., Markiewicz, W.J., Keller, H.U. 1999. Mass and Energy Balance in the Near-Surface Layers of a Cometary Nucleus. Icarus 140, 173-188.

Skorov, Y.V., van Lieshout, R., Blum, J., Keller, H.U. 2011. Activity of comets: Gas transport in the near-surface porous layers of a cometary nucleus. Icarus, 212, 867-876.

Steiner, G. 1990. Two considerations concerning the free molecular flow of gases in porous ices. Astronomy and Astrophysics 240, 533-536.

Sunshine, J. M.; A'Hearn, M. F.; Groussin, O.; Li, J.Y.; Belton, M. J. S.; Delamere, W. A.; Kissel, J.; Klaasen, K. P.; McFadden, L. A.; Meech, K. J.; Melosh, H. J.; Schultz, P. H.; Thomas, P. C.; Veverka, J.; Yeomans, D. K.; Busko, I.; Desnoyer, M.; Farnham, T. L.; Feaga, L. M.; Hampton, D.; Lindler, D.;Wellnitz, D. D. 2006. Exposed water ice deposits on the surface of Comet Tempel 1. Science 311, 1453–1455.

Toth, I.; Lisse, C. M. 2006. On the rotational breakup of cometary nuclei and centaurs, Icarus, 181, 1, 162-177.

Weidling, R., Güttler, C., Blum, J., & Brauer, F. 2009. The Physics of Protoplanetesimal Dust Agglomerates. III. Compaction in Multiple Collisions. Astrophysical Journal, 696, 2036 -2043.

Weidling et al. 2011. Free Collisions in a Microgravity Many-Particle Experiment. I. Dust Aggregate Sticking at Low Velocities, accepted by Icarus (arXiv:1105.3909).

Wellnitz, D. D.; White, R. L. 2005. Deep Impact: Excavating Comet Tempel 1, Science, 310, 5746, 258-264.

Wilner, D.J., D'Alessio, P., Calvet, N., Claussen, M.J., Hartmann, L. 2005. Toward Planetesimals in the Disk around TW Hydrae: 3.5 Centimeter Dust Emission. The Astrophysical Journal 626, L109-L112.

Zolensky, M. E.; Zega, T. J.; Yano, H.; Wirick, S.; Westphal, A. J.; Weisberg, M. K.; Weber, I.; Warren, J. L.; Velbel, M. A.; Tsuchiyama, A.; Tsou, P.; Toppani, A.; Tomioka, N.; Tomeoka, K.; Teslich, N.; Taheri, M.; Susini, J.; Stroud, R.; Stephan, T.; Stadermann, F. J.; Snead, C. J.; Simon, S.n B.; Simionovici, A.; See, T. H.; Robert, F.; Rietmeijer, F. J. M.; Rao, W.; Perronnet, M. C.; Papanastassiou, D. A.; Okudaira, K.; Ohsumi, K.; Ohnishi, I.; Nakamura-Messenger, K.; Nakamura, T.; Mostefaoui, S.; Mikouchi, T.; Meibom, A.; Matrajt, G.; Marcus, M. A.; Leroux, H.; Lemelle, L.; Le, L.; Lanzirotti, A.; Langenhorst, F.; Krot, A. N.; Keller, L. P.; Kearsley, A. T.; Joswiak, D.d; Jacob, D.; Ishii, H.; Harvey, R.; Hagiya, K.; Grossman, L.e; Grossman, J. N.; Graham, G. A.; Gounelle, M.; Gillet, P.; Genge, M. J.; Flynn, G.; Ferroir, T.; Fallon, S.; Ebel, D. S.; Dai, Z. R.; Cordier, P.; Clark, B.; Chi, M.; Butterworth, A. L.; Brownlee, D. E.; Bridges, J. C.; Brennan, S.; Brearley, A.; Bradley, J. P.; Bleuet, P.; Bland, P. A.; Bastien, R. 2006. Mineralogy and Petrology of Comet 81P/Wild 2 Nucleus Samples. Science 314, 1735-1739.

Zsom, A., Ormel, C. W., Güttler, C., Blum, J., & Dullemond, C. P. 2010. Astronomy and Astrophysics, 513, A57.

Zsom, A., Ormel, C. W., Dullemond, C. P., & Henning, T. 2011. arXiv:1107.5198.

**Figure captions.**

Figure 1. (a) Schematic sketch of the formation scenario of cometesimals used in our comet nucleus model. Dust and ice aggregates are captured by the growing cometesimal due to a gravitational instability so that they hit the surface at about escape speed. As the threshold velocity for direct sticking is well below the escape speed, the aggregates will bounce a few times before they stick roughly at the sticking threshold velocity. (b) Sketch of the two-layer model of comet nuclei. The bulk of the comet nucleus consists of dust and ice aggregates, which themselves consist of µm-sized dust and ice grains. The ice free surface layer in the case shown is more porous than the rest of the nucleus. An alternative model assumes restructuring of the dust aggregates in the ice-free surface layer until a denser configuration is reached (not shown).

Figure 2. Top panels: Temperature of the sublimating $H_2O$ ice surface as a function of the thickness of the dust layer for several sizes of dust aggregates (from $10^{-5}$ m to $10^{-2}$ m, as indicated in the Figure legend) for two heliocentric distances (at the left column $R = 1$ AU, at the right column $R = 3$ AU) and different permeability of the dust layer (for $\phi_p = 0.3$ curves marked by circle). Bottom panels: The corresponding gas pressure underneath the porous dust crust.

Figure 3. Top panels: Temperature of the sublimating $CO_2$ ice surface as a function of the thickness of the dust layer for several sizes of dust aggregates (from $10^{-5}$ m to $10^{-2}$ m, as indicated in the Figure legend) for two heliocentric distances (at the left column $R = 1$ AU, at the right column $R = 3$ AU) and different permeability of the dust layer (for $\phi_p = 0.3$ curves marked by circle). Bottom panels: The corresponding gas pressure underneath the porous dust crust.

Figure 4. Comparison of the strength of the dust crust calculated for two filling factors (bold dashed curves) and $H_2O$ gas pressure calculated for different dust-

aggregate sizes (from $10^{-5}$ m to $10^{-2}$ m). Different solid curves correspond to different thicknesses of the dust crust (triangles for $L=10^{-3}$ m, diamonds for $L=10^{-2}$ m, squares for $L=3\cdot10^{-2}$ m, and circles for $L=5\cdot10^{-2}$ m) and different permeability of the dust layer (solid curves for $\phi_p = 0.6$, and dashed curves for $\phi_p = 0.3$). Model results for $R = 1$ AU (panel $a$) and $R = 3$ AU (panel $b$) are presented.

Figure 5. Comparison of the strength of the dust crust calculated for two filling factors (bold dashed curves) and $CO_2$ gas pressure calculated for different dust-aggregate sizes (from $10^{-5}$ m to $10^{-2}$ m). Different solid curves correspond to different thicknesses of the dust crust (triangles for $L=10^{-3}$ m, diamonds for $L=10^{-2}$ m, squares for $L=3\cdot10^{-2}$ m, and circles for $L=5\cdot10^{-2}$ m) and different permeability of the dust layer (solid curves for $\phi_p = 0.6$, and dashed curves for $\phi_p = 0.3$). Model results for $R = 1$ AU (panel $a$) and $R = 3$ AU (panel $b$) are presented.

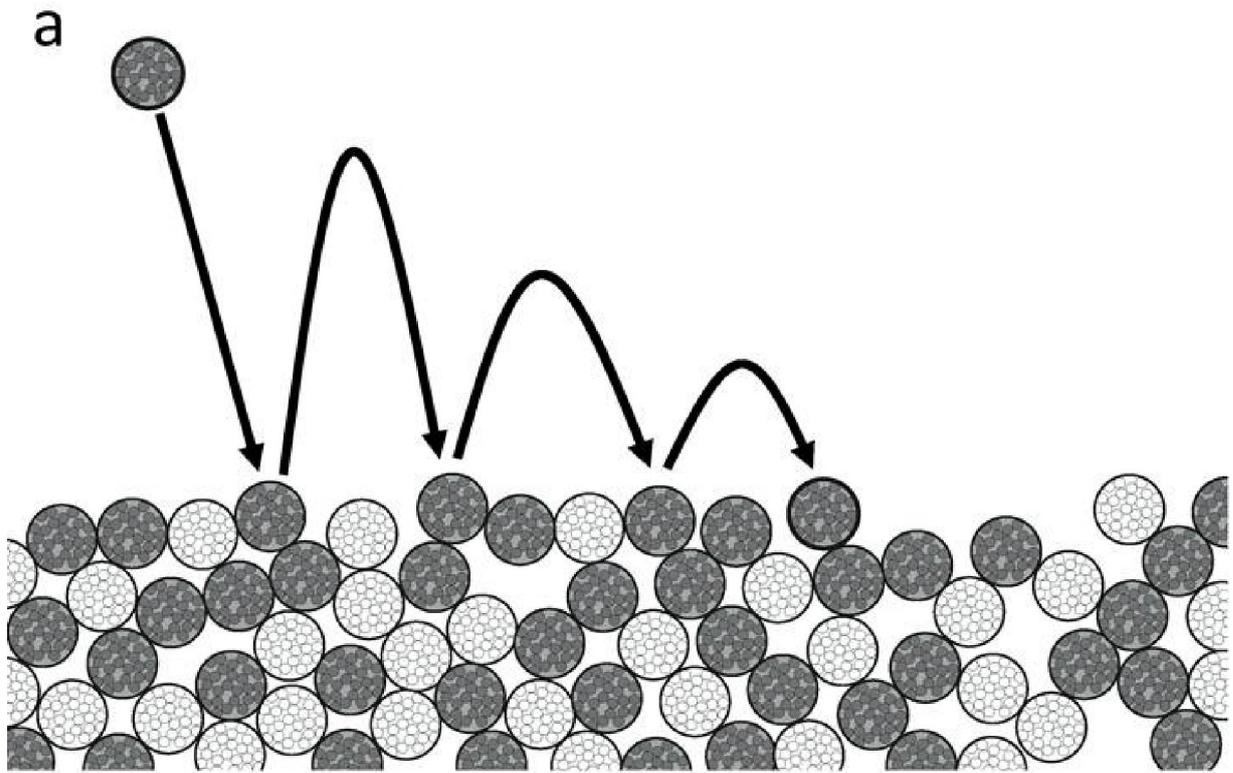

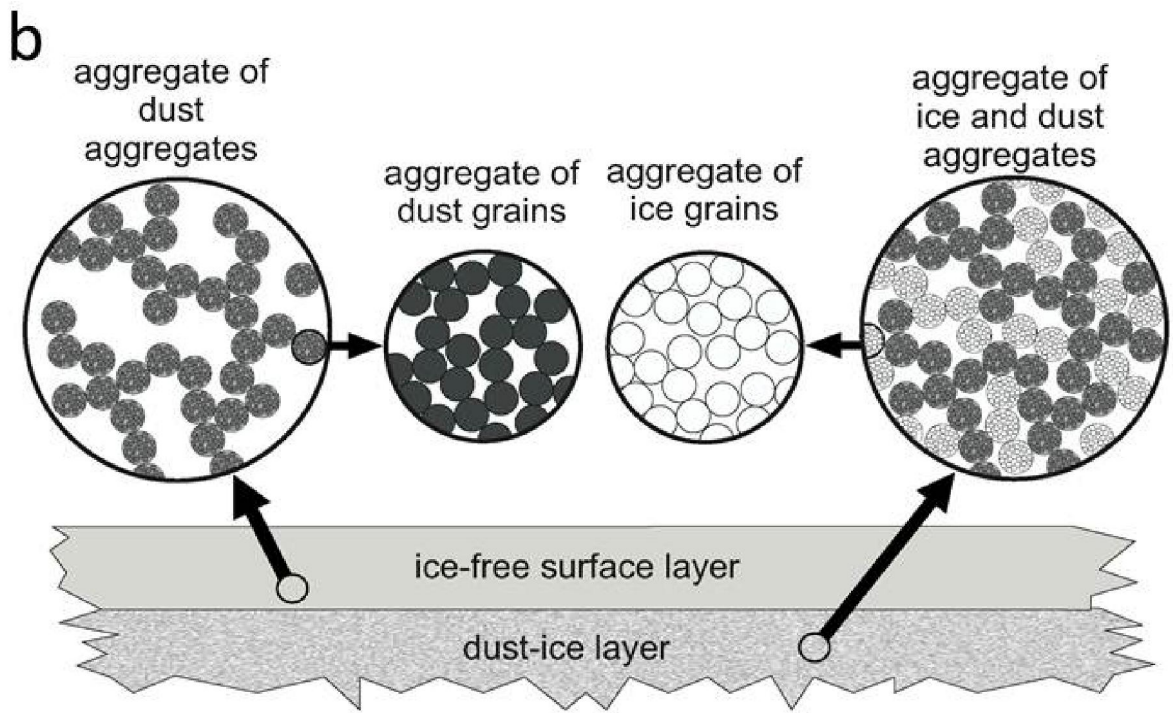

Fig. 1

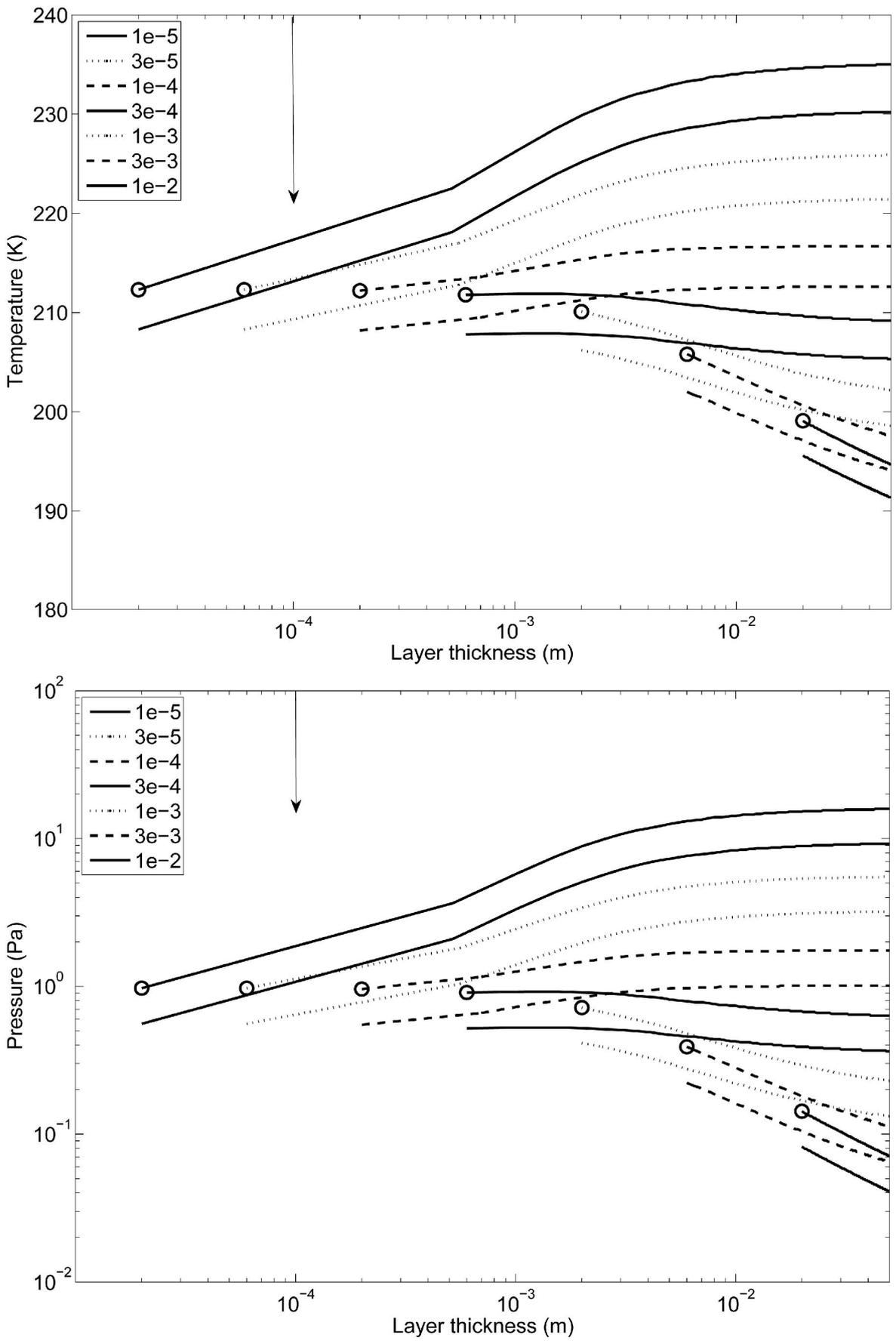

Fig. 2a

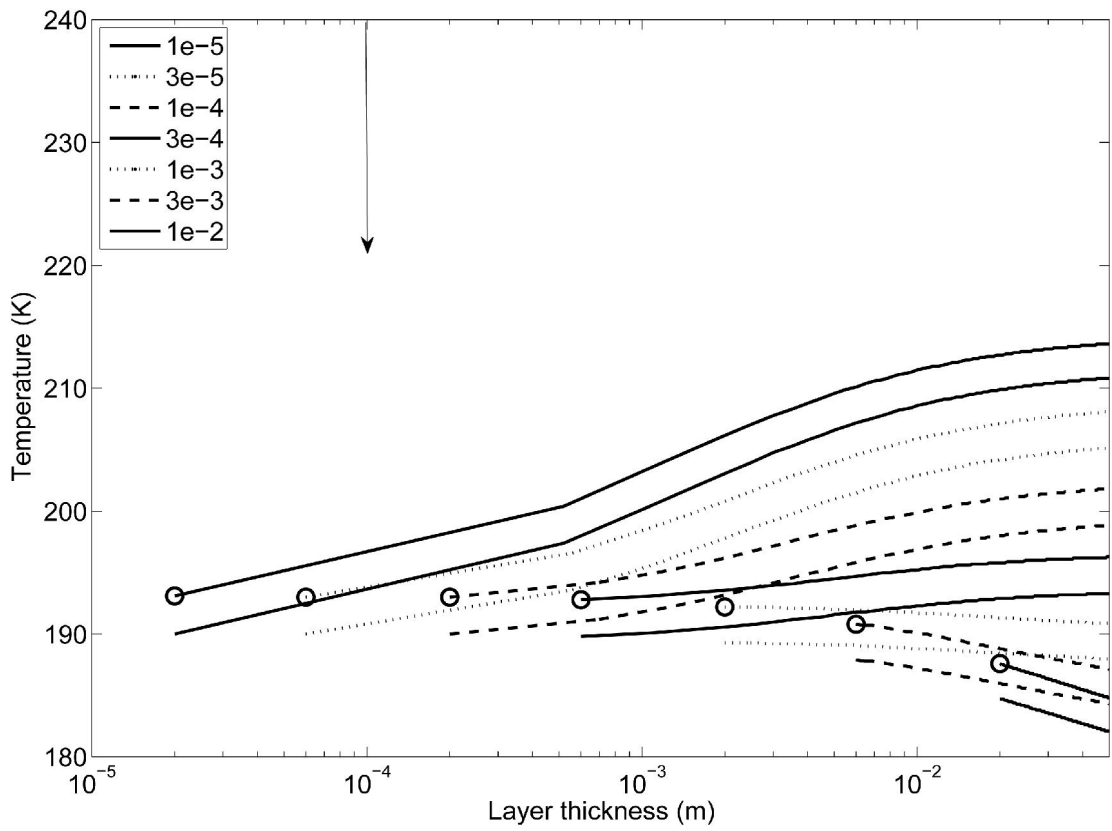

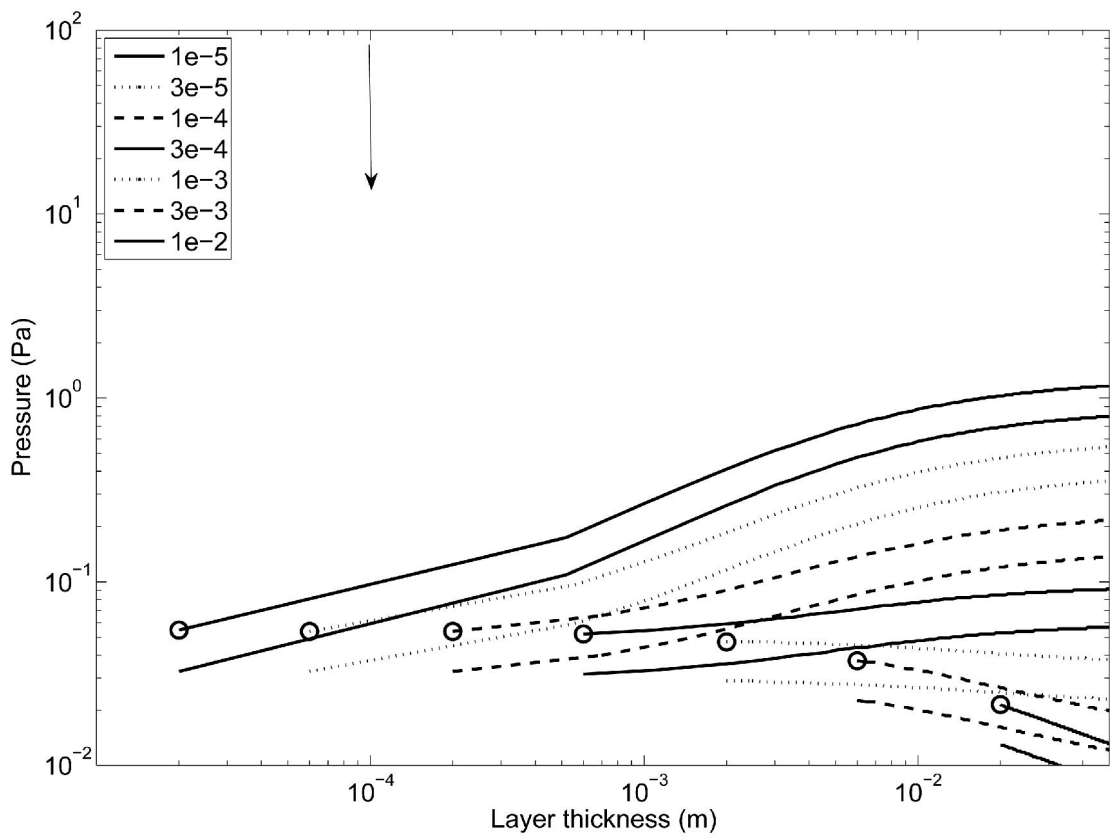

Fig. 2b

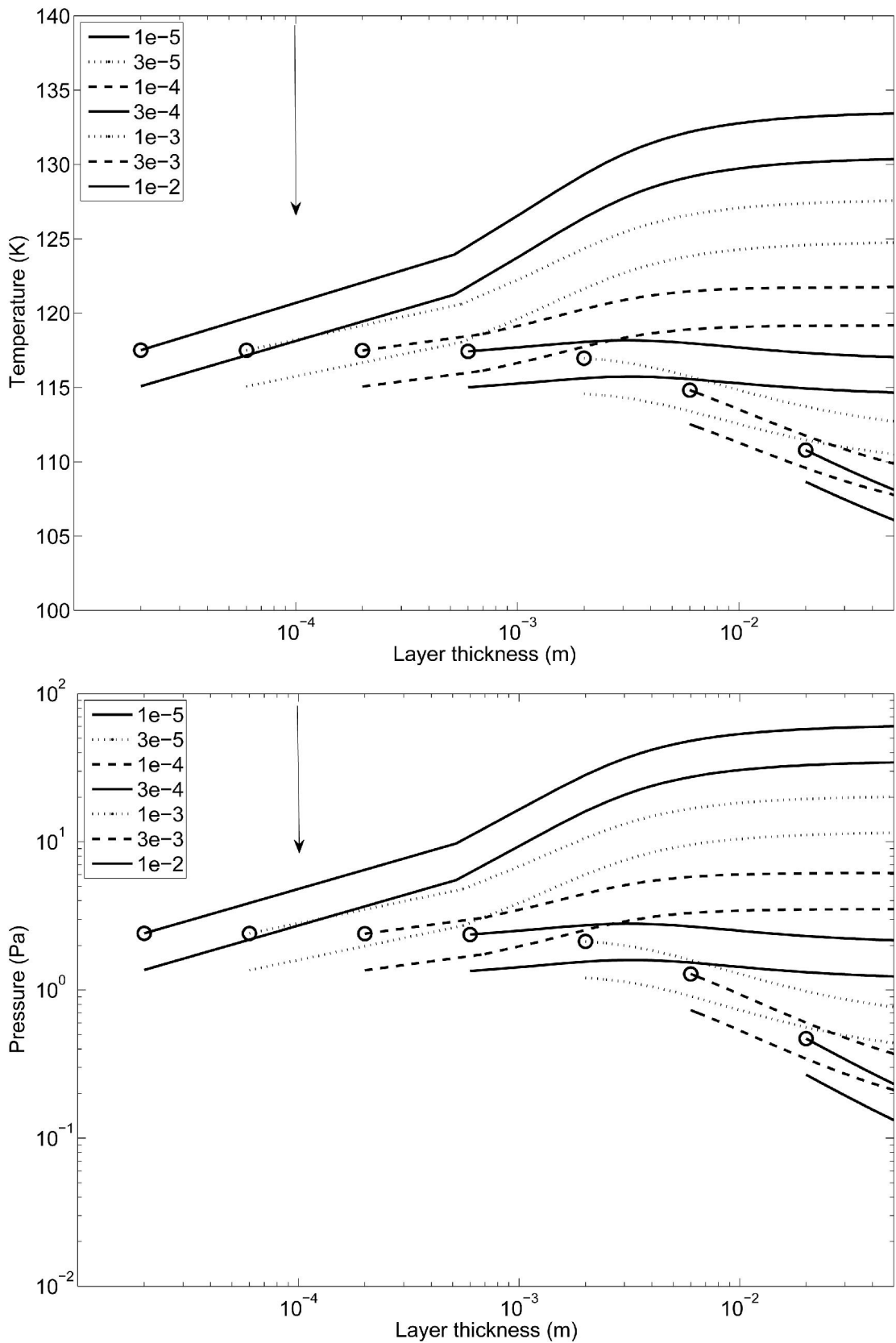

Fig. 3a

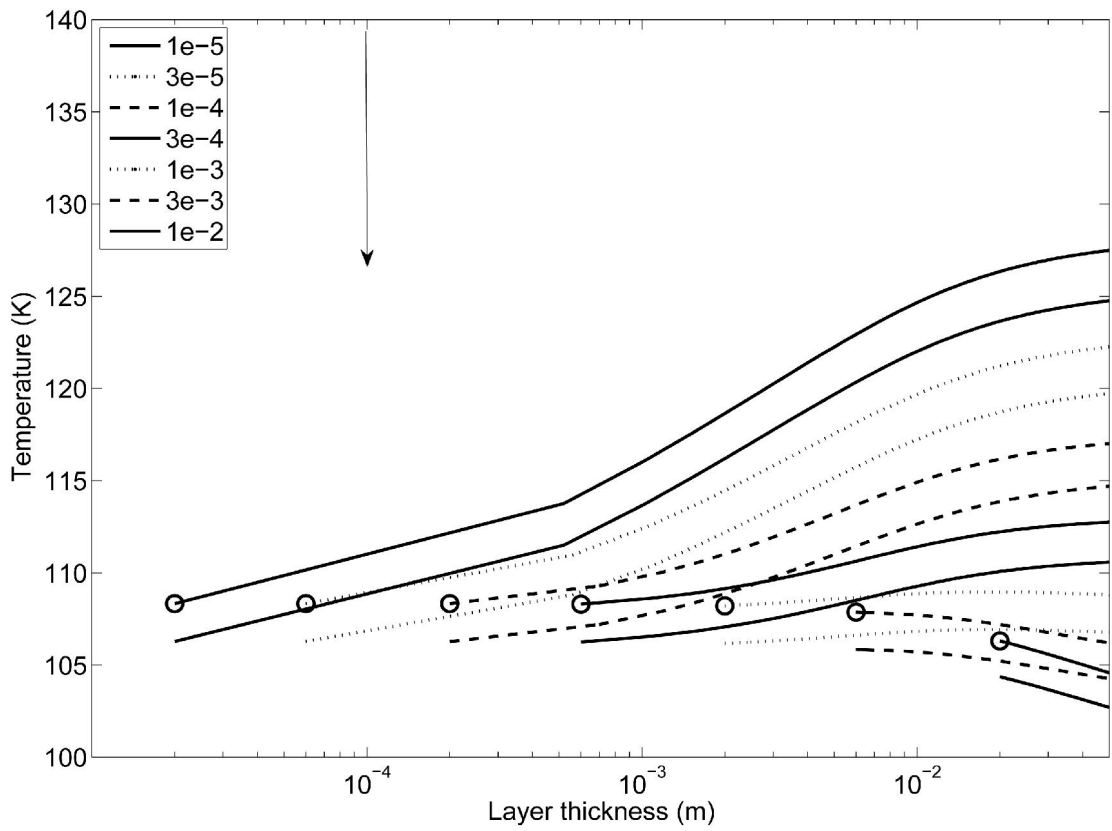

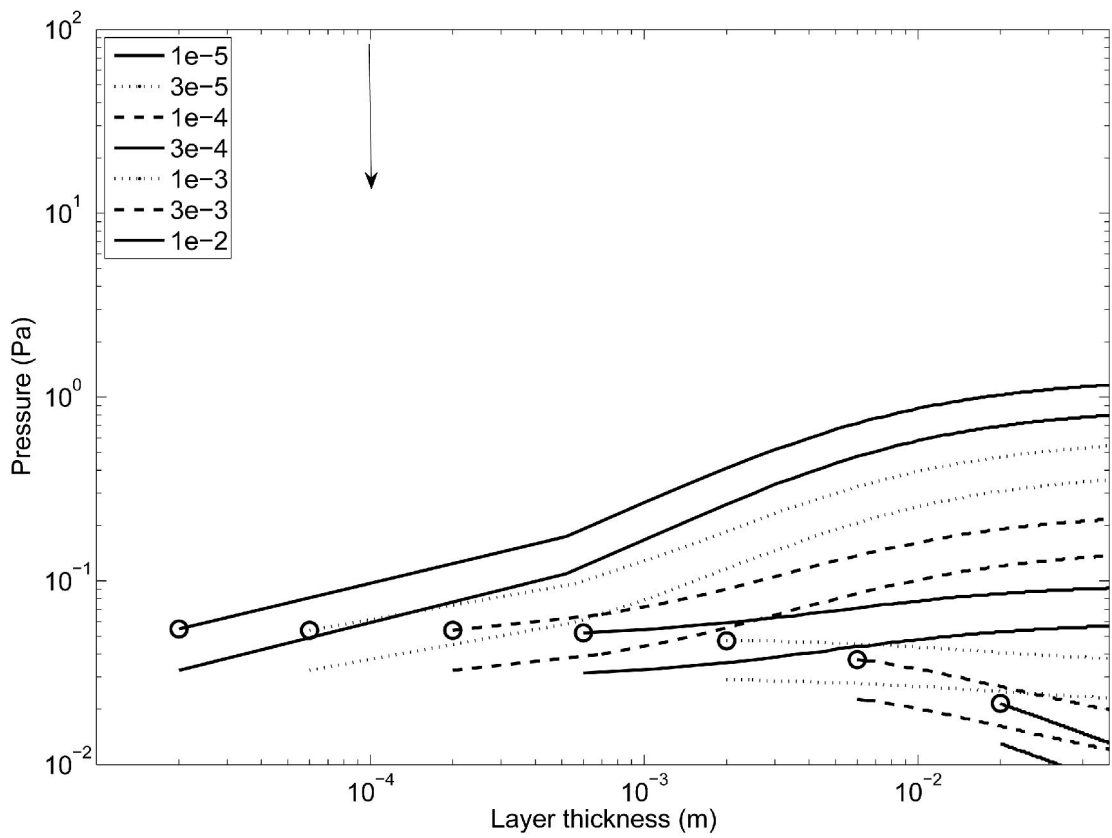

Fig. 3b

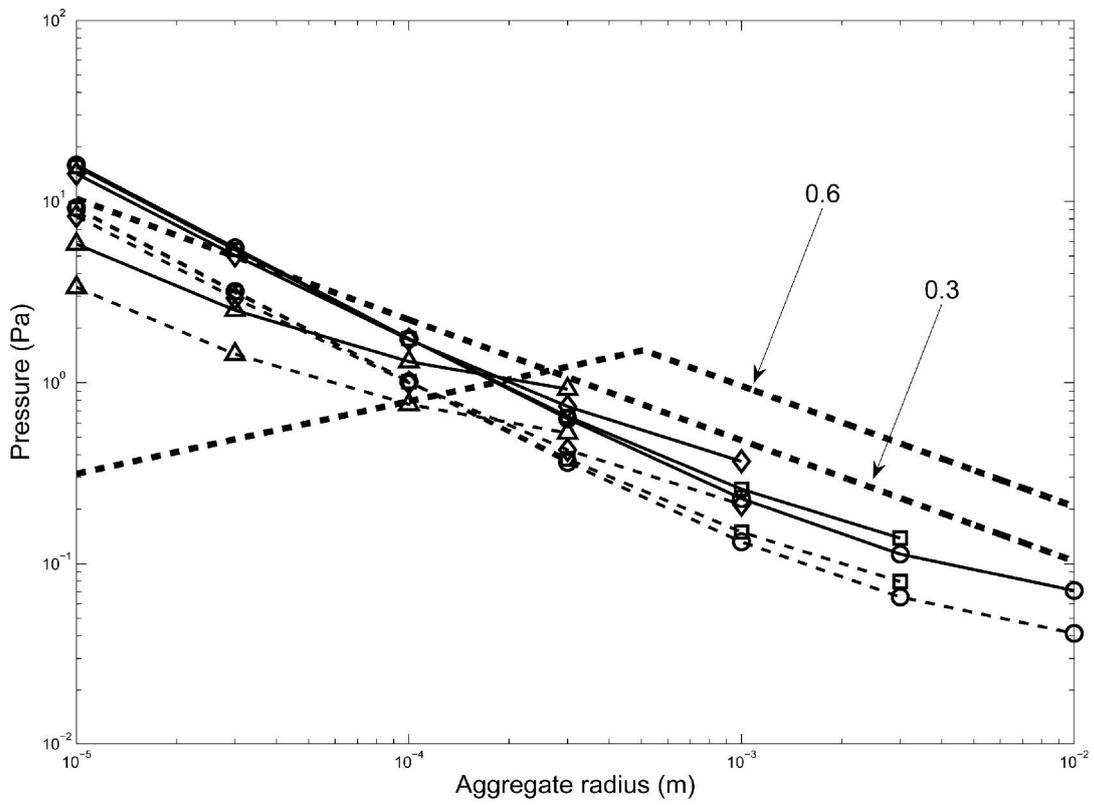

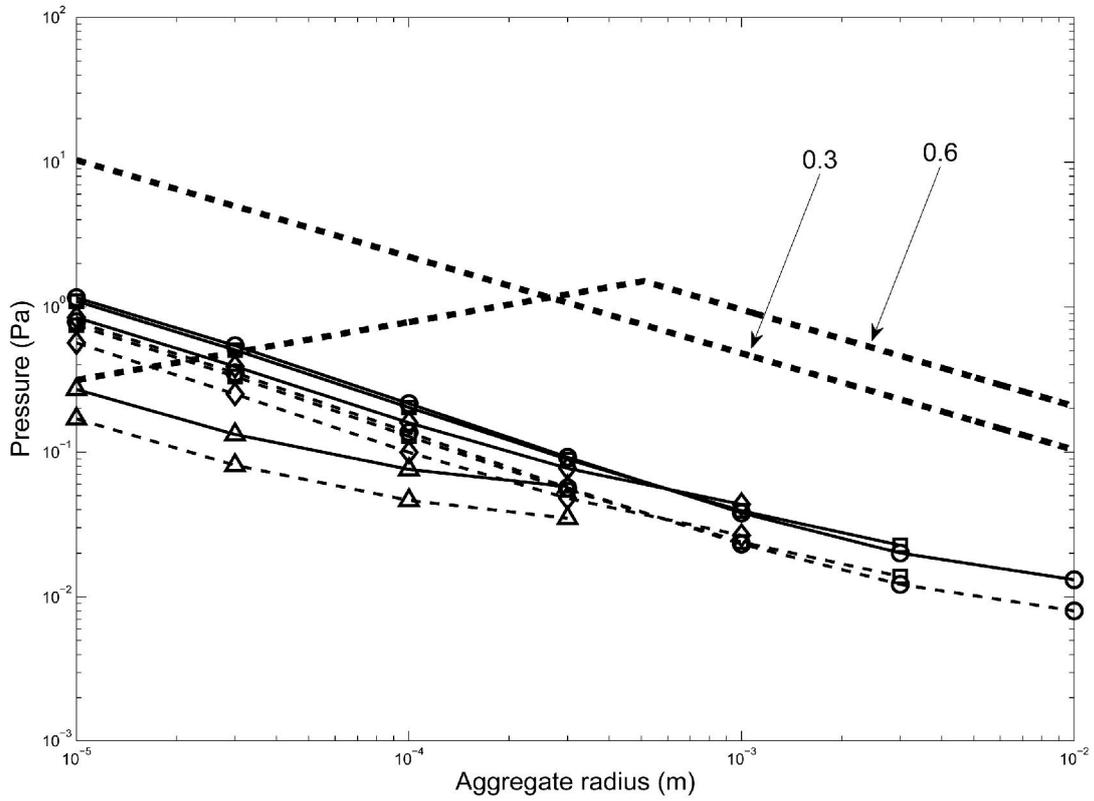

Fig. 4

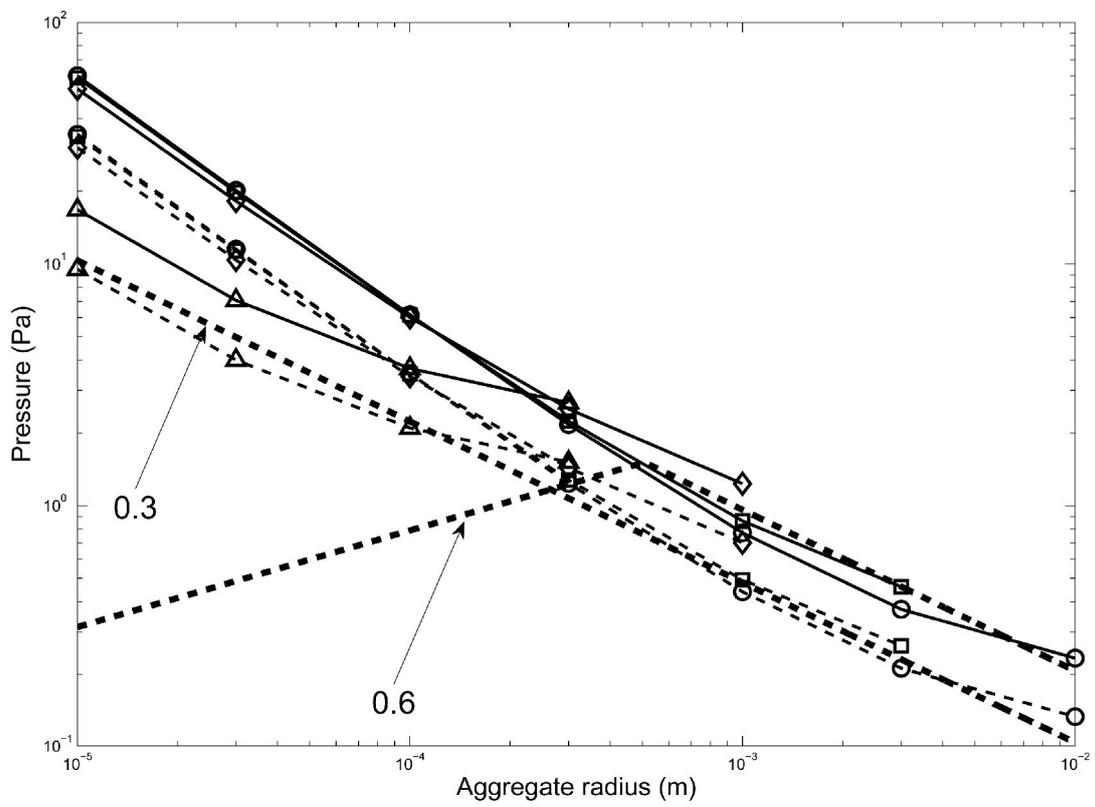

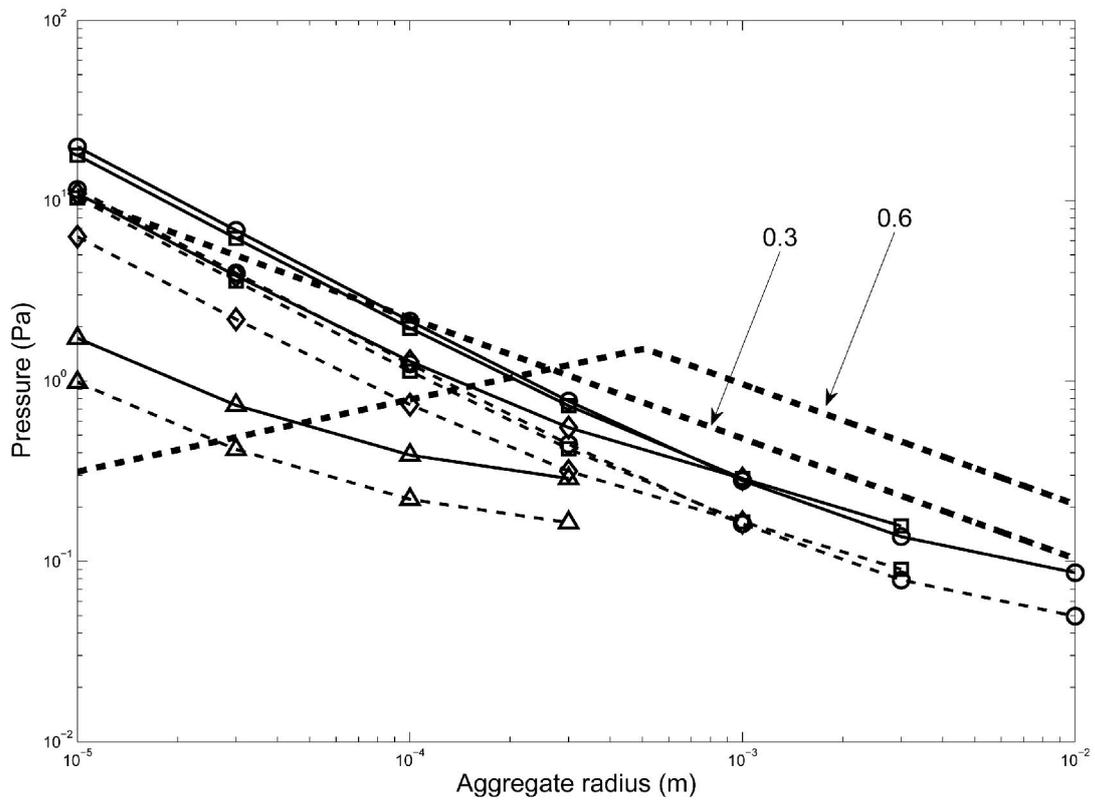

Fig. 5